# Enhancing thermal conductivity of bulk polyethylene along two directions by paved crosswise laminate

Xiaoxiang Yu[1,2,#], Chengcheng Deng[3,#], Xiaoming Huang[3,*], Nuo Yang[1,2,*]

[1] State Key Laboratory of Coal Combustion, Huazhong University of Science and Technology, Wuhan 430074, P. R. China

[2] Nano Interface Center for Energy (NICE), School of Energy and Power Engineering, Huazhong University of Science and Technology, Wuhan 430074, P. R. China

[3] School of Energy and Power Engineering, Huazhong University of Science and Technology, Wuhan 430074, P. R. China

# X. Y. and C. D. contributed equally to this work.

* To whom correspondence should be addressed. E-mail: nuo@hust.edu.cn (NY), xmhuang@hust.edu.cn (XH)

## Abstract

Recently, some reports show that the ultra-low thermal conductivity of bulk polymers can be enhanced along one direction, which limits its applications. Here, we proposed paved crosswise laminate methods which can enhance the thermal conductivity of bulk polyethylene (PE) along two directions. We find that the thermal conductivity of paved crosswise polyethylene laminate (PPEL) reaches as high as 181 W/m-K along two in-plane directions, which is three orders of magnitude larger than bulk amorphous polyethylene and even more than two times larger than PE single chain (54 W/m-K). The analyses of mechanism indicated that PPEL is a much more crystal-like structure due to the inter-chain van der Waals interactions. Our study may provide guides on the design and fabrication of polymer structures with high thermal conductivity.

# Introduction

The polymers play indispensable roles in our daily life due to their outstanding physical and chemical properties. However, they are seldom used for heat removing because of its poor thermal properties. The bulk amorphous polyethylene (PE), one of the simplest and most widely used materials, is often considered as thermal insulator with an ultra-low thermal conductivity ($\kappa_{bulk}$) as ~ 0.1 W/m-K.[1-3]

Recently, one-dimensional single PE chain (SPEC) was found to have a high thermal conductivity as 350 W/m-K at room temperature by molecular dynamics (MD) simulations.[4,5] The largest contributions of thermal conductivity come from the longitudinal mode, which have a very high group velocity as ~ 16000 m/s.[4] There are some factors affecting thermal conductivities of polymer chains, such as morphology[6-8] and length[9] etc. In addition, experimental measurements also showed that polymer nanofibers have outstanding thermal conductivities[10], such as the thermal conductivities of high-quality ultra-drawn PE nanofibers as 104 W/m-K[11] and chain-oriented polythiophene nanofibers as 43.3 W/m-K.[12]

Besides one-dimensional (1D) PE chains and fibers, some efforts on fabricating applicable bulk materials have been made recently. Highly aligned PE films[13,14] and PE nanowire arrays[15,16] have high thermal conductivities. Furthermore, some guides were provided to improve thermal properties, e.g. by doping carbon nanotubes. Our previous studies showed that the thermal conductivity of aligned carbon nanotube-PE composites can reach 99.5 W/m-K simulated using MD.[17] However, in previous works, thermal transport has been enhanced just along one direction, which limits their applications.

So, how to enhance the thermal conductivity of PE chains along more than one direction? Here, we raise a new idea on enhancing the thermal conductivity of bulk polymers along two directions. That is, we propose a feasible bulk structure as paved crosswise polyethylene laminate (PPEL) (as shown in Fig. 1). Moreover, our idea is viable and applicable, because there are already analogous electrospinning thin films[18] and three-dimensional printing structures[19].

In this paper, we investigated numerically the thermal conductivities of PPEL. We firstly showed a description of model and simulation details. Secondly, the along-chain thermal conductivities of SPEC ($\kappa_{SPEC}$) and PPEL ($\kappa_{PPEL}$) were calculated. Lastly, we analyzed the mechanisms of the high thermal transport in PPEL.

## Model and simulation details

Figure 1 depicts the structure of a simulation cell of PPEL. The aligned chains are paved layer by layer. And the aligned directions are crosswise for two adjacent layers. All simulations are performed by the Large-scale Atomic/Molecular Massively Parallel Simulator (LAMMPS) package developed by Sandia National Laboratories.[20] In simulations, the periodic boundary conditions are applied in all three directions. The interatomic interactions are described by an adaptive intermolecular reactive empirical bond order (AIREBO) potential,[21,22] which is developed from the second-generation Brenner potential.[23]

The equilibrium molecular dynamics (EMD) method, named as Green-Kubo method, is employed to calculate the thermal conductivity, which is widely used in calculating thermal conductivity of bulk structures.[24,25] The Green-Kubo formula[26,27] relates the ensemble average of the heat current ($\vec{J}$) auto-correlation to the thermal conductivity ($\kappa$). The heat current is defined as

$$\vec{J} = \frac{1}{V}\left[\sum_i e_i \vec{v}_i + \frac{1}{2}\sum_{i<j}\left(\vec{f}_{ij} \bullet \left(\vec{v}_i + \vec{v}_j\right)\right)\vec{x}_{ij}\right] \quad (1).$$

The thermal conductivity is derived from the Green-Kubo equation as

$$\kappa = \frac{V}{3k_B T^2}\int_0^{\tau_0}\left\langle \vec{J}(0) \bullet \vec{J}(\tau) \right\rangle d\tau \quad (2)$$

where $k_B$ is the Boltzmann constant, $V$ is the system volume, $T$ is the temperature, $\tau$ is the correlation time, $\tau_0$ is the integral upper limit of heat current auto-correlation function (HCACF), and the angular bracket denotes an ensemble average.

Generally, the temperature in MD simulation, $T_{MD}$, is calculated by the formula

$$\langle E \rangle = \sum_{1}^{N} \frac{1}{2} m_i v_i^2 = \frac{3}{2} N k_B T_{MD} \quad (3)$$

where $E$ is total kinetic energy of the group of atoms, and $N$ is number of total atoms. This equation is valid at high temperature.

The velocity Verlet algorithm[28] is employed to integrate equations of motion. The time step is set as 0.1 fs. The structures are minimized by standard conjugate-gradient energy-minimization methods in LAMMPS. After relaxing the system in isobaric-isothermal ensemble (*NPT*) by 100 ps, it runs more 100 ps in microcanonical ensemble (*NVE*). Then, it runs another 5 ns in *NVE* for recording the heat current at each 1 fs and calculating the thermal conductivity using Eq. 2. We use a combination of time and ensemble sampling to obtain a better average statistics. The result represents the average of 5 independent simulations with different initial conditions.

## Results and Discussions

Compared with the thermal conductivity along two in-plane directions (x and y), the thermal conductivity of PPEL along the cross-plane direction (z) is much lower, as ~ 0.1 W/m-K by our calculations. So we will focus on thermal conductivity along two in-plane directions.

The main results are shown in Fig. 2. The temperature dependences of thermal conductivity ($\kappa$) are obtained by EMD simulations. The value of $\kappa_{SPEC}$ reaches 54 W/m-K at room temperature, which is close to previous results as 61 W/m-K[6] and 57 W/m-K[17]. Surprisingly, the thermal conductivity of PPEL along two in-plane directions (x and y) achieves as high as 181 W/m-K at room temperature. Firstly, PPEL possesses good thermal properties along two directions, which goes a step further in the enhancement of thermal property in polymers and extends the scope of applications. Moreover, the enhancement of the value of thermal conductivity is outstanding. That is, $\kappa_{PPEL}$ is not only three orders of magnitude larger than that of bulk PE ($\kappa_{bulk}$), but also more than two times larger than that of single PE chain ($\kappa_{SPEC}$). Besides, the thermal conductivity of PPEL is much larger than that of most of PE composites, such as PE nanofibers (104 W/m-K),[11] PE nanowire array (21.1 W/m-K),[15] aligned carbon nanotube-PE composites (99.5 W/m-K).[17]

What's more, the value of thermal conductivity of PPEL, such as 181

W/m-K at 300K, is the effective thermal conductivity, which use the whole cross section area and is similar to the measurement results. When we analyze the contribution to the thermal transfer, the layers, where PE chains are perpendicular to direction of heat transfer, has an ultra-low thermal conductivity. So, most of the contribution comes from the layers where PE chains are parallel to direction of heat transfer, whose thermal conductivity is estimated to twice of the effective thermal conductivity, such as ~ 362 W/m-K at 300 K. Generally, in the previous works,[15,16,22] the thermal conductivity of 1D PE chain is much larger than that of amorphous bulk PE or other bulk composites, due to the largely reduction of inter-chain phonon scatterings in 1D chain. Interesting, our results of PPEL show that the thermal conductivity of bulk PPEL is much larger than $\kappa_{SPEC}$.

The enhancement of thermal conductivity in bulk paved crosswise polyethylene laminate attributes to two mechanisms. Firstly, comparing with randomly distributed in bulk amorphous PE, chains are aligned and paved into laminate. Similar to the structure of aligned carbon nanotube-polyethylene composites,[17] nanofiber,[11] and nanowire array,[15,16] the alignment weakens the wiggles of chains and reduces the inter-chain scatterings, which ensures a good phonon transport along chains. Secondly, the crosswise layout not only endows the good thermal transport along two directions but is beneficial for the enhancement of the thermal transport. The nonbonding interactions, van der Waals forces, between orthogonal chains have a weak confinement on the atomic vibrations and a reduction of segmental rotations. That is, the atomic vibrations are more crystal-like which will increase the thermal conductivity further.

Next, the mechanism of phonon transport in PPEL structure is further analyzed. As reported in previous works,[6,7,29] the wiggles and segmental rotations will lead to phonon-phonon scattering and suppress phonon transport in PE chains. To quantify lattice orders of PE chains, we calculated the radial distribution functions (RDFs) and the probability distribution of the dihedral angle of the C-C-C-C backbone. Figure 3(a) shows the significant differences of RDFs between SPEC and PPEL. For SPEC, there is no obvious peak in RDFs. Correspondingly, the view along chain axis (the inset of blue dots) indicates that atomic vibrations in SPEC are full of wiggles. By contrast, there are two sharp peaks in RDFs of a PE chain in PPEL. In correspondence with the view along chain axis (the inset of red dots), the two peaks in RDFs of PPEL correspond to the equilibrium positions where C atoms and H atoms are located. It denotes that PPEL has much less wiggles than SPEC, which is conducive to a better phonon transporting phonons and a higher thermal conductivity.

Besides, it is shown in Fig. 3(b) that the probability distributions of the dihedral angles of the backbone. It demonstrates that there is a much steeper peak for a chain in PPEL than the SPEC. Correspondingly, there are much weaker segmental rotations of a PE chain in PPEL than the SPEC. That is, the PPEL is almost crystalline because of slight wiggles and weak segmental rotations, which contribute to its high along-chain thermal conductivity.

# Conclusions

We proposed a new idea to enhance the thermal conductivity of bulk polyethylene by paved crosswise laminate methods. The molecular dynamics simulation results show that paved crosswise polyethylene laminate has a high thermal conductivity as 181 W/m-K along two in-plane directions, which is more than three orders of magnitude higher than bulk amorphous PE and even three times higher than the single PE chain. By analyzing the radial distribution functions and probability distributions of dihedral angles of backbone, it is found that PPEL has a much better crystal-like lattice vibration than the SPEC. The crosswise layout confines the wiggles and segmental rotations via inter-chain van der Waals interactions. Besides, the crosswise layout also endows good thermal transport along two directions.

The results may provide a significant guide on the design and fabrication of polymer structure with high thermal conductivity. The idea would also be good for other one-dimensional materials e.g. nanotube and nanowire. Besides crosswise layout, it would be generalized to multidirectional layout structures, such as hexagonal and octagonal conformations.

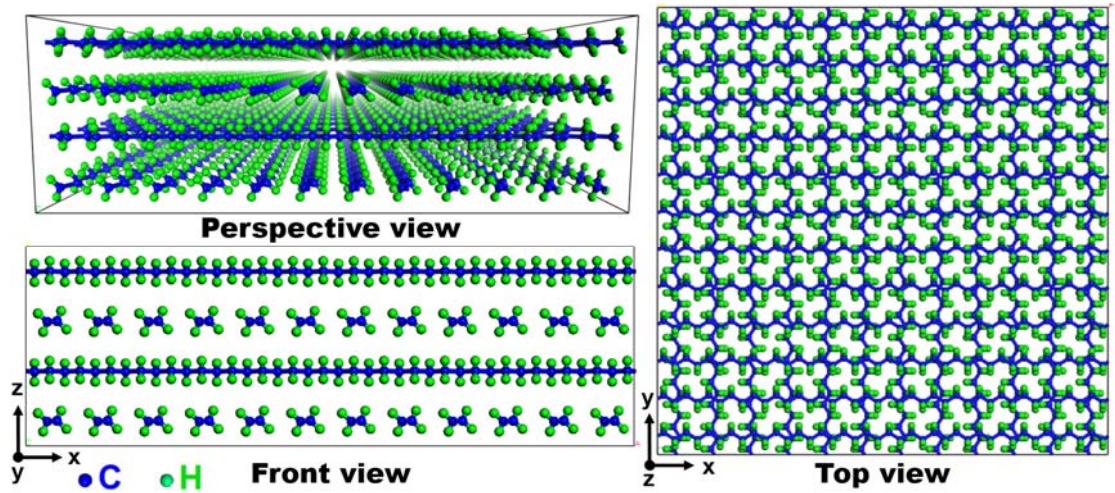

Figure 1. Schematic views of a simulation cell of paved crosswise polyethylene laminate (PPEL). In simulations, periodic boundary conditions are applied in all three directions. The aligned chains are paved layer by layer. And the aligned directions are crosswise for two adjacent layers. We focus on the thermal conductivity along two in-plane directions (x and y).

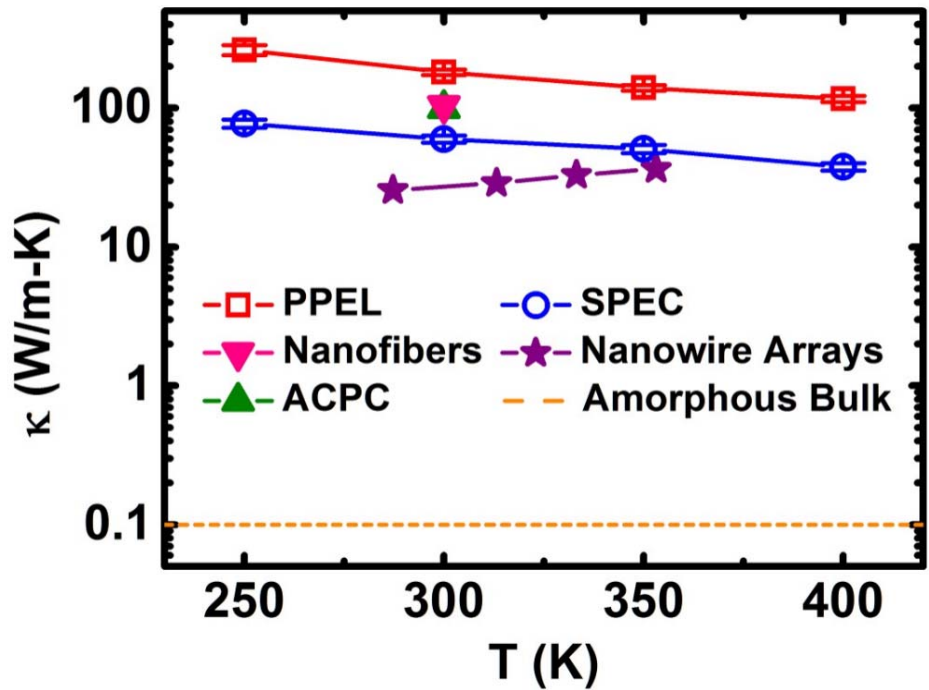

Figure 2. The temperature dependence of thermal conductivity (κ) of PPEL. The PPEL has the highest values of κ comparing with the results of SPEC, PE nanofibers,[11] aligned carbon nanotube-PE composites (ACPC),[17] PE nanowire arrays,[15] and amorphous bulk PE.[1] At 300 K, $\kappa_{PPEL}$ (181 W/m-K) is not only three orders of magnitude larger than $\kappa_{bulk}$ (0.1 W/m-K), but more than two times larger than $\kappa_{SPEC}$ (54 W/m-K). $\kappa_{bulk}$ at different temperature are simply considered the same.[1]

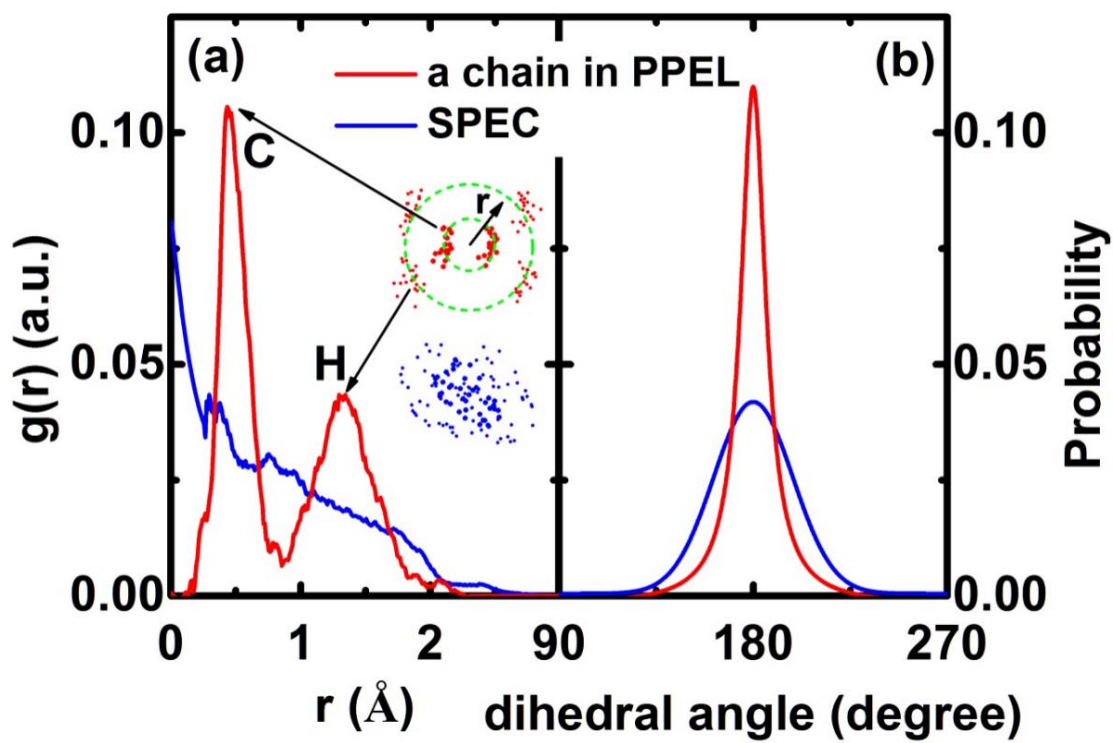

Figure 3. (a) The radial distribution functions (RDFs) g(r) for a PE chain in PPEL and SPEC. There is no obvious peak in RDFs of SPEC. Correspondingly, the view along chain axis (the inset of blue dots) indicates that atomic vibrations in SPEC are full of wiggles. By contrast, there are two sharp peaks in RDFs of a PE chain in PPEL. In correspondence with the view along chain axis (the inset of red dots), the two peaks in RDFs of PPEL correspond to the equilibrium positions (green dashed circles) where C atoms and H atoms are located. (b) The probability distributions of the dihedral angles of C-C-C-C backbone. There is a much steeper peak for a chain in PPEL than the SPEC. That is,

there are much weaker segmental rotations of a PE chain in PPEL than the SPEC.

# Acknowledgements

The work was supported by the National Natural Science Foundation of China No. 51576076 (N. Y.) and No. 51576076 (X. H.), and the Self-Innovation Foundation of HUST No. 2015QN037 (C. D.). The authors acknowledge stimulating discussions with Quanwen Liao and Meng An. The authors thank the National Supercomputing Center in Tianjin (NSCC-TJ) and the High Performance Computing Center Experimental Testbed in SCTS/CGCL for providing help in computations.

# Supporting Information

1. Size dependence of thermal conductivity of single PE chain (SPEC)

In order to get the appropriate system length for EMD simulations, the size dependence of thermal conductivity of SPEC at different temperatures (250 ~ 400 K) is presented in Fig. S1. For one thing, there is a negative temperature dependence of $\kappa_{SPEC}$. It is the reason that chain twists more strongly and more anharmonic phonon scattering appears at higher temperature. For another, $\kappa_{SPEC}$ remains constant with increasing simulation length. Therefore, 5 nm are selected.

2. Relaxation in *NPT* ensemble

To get the relaxed PPEL structure, initial structure is minimized by standard conjugate-gradient energy-minimization methods in LAMMPS. Then, the system runs in isobaric-isothermal ensemble (*NPT*) by 100 ps. Figure S2 shows the stress as a function of time, indicating no stress in all three directions after *NPT* process.

3. Radial distribution function (RDF)

The reference atom ($x_0$, $y_0$, $z_0$) is the average coordinates of all atoms. The chain is along $z$ direction. RDF is calculated by $g(r) = N/(2n-1)$. The distance ($r$) is defined as $r = \sqrt{(x-x_0)^2 + (y-y_0)^2}$. $N$ is the count of atoms in $n$th annulus ( $R \leq r < R+d$ ). $A(r)$ is the area of annulus. The width of each annulus $d$ is set to 0.2 Å. The value of $\pi d^2$ is normalized as 1, thus the area of $n$th annulus is $A(r) = \pi(nd)^2 - \pi((n-1)d)^2 = (2n-1)\pi d^2 = (2n-1)$.

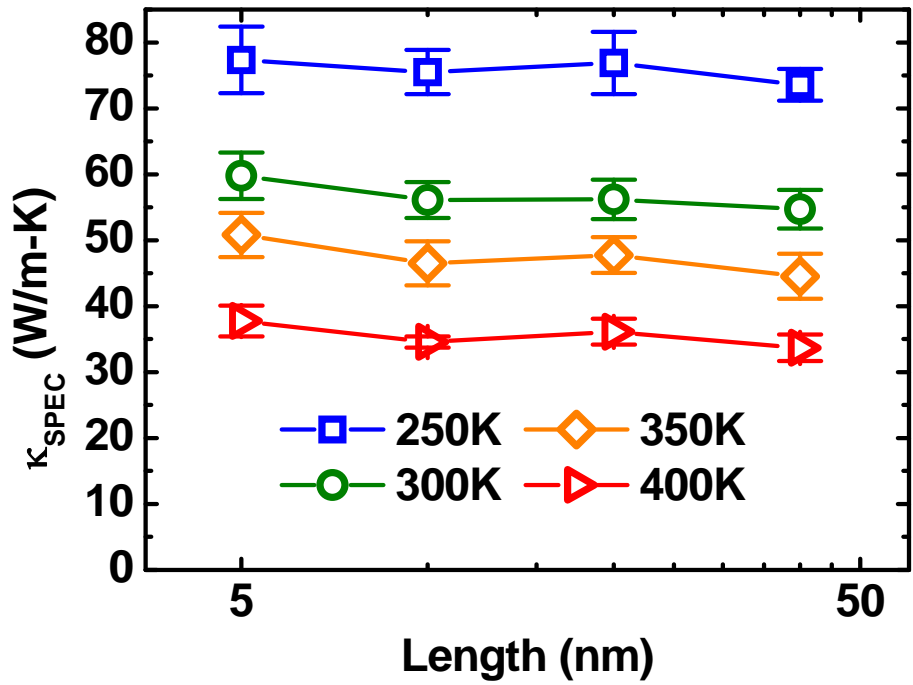

Figure S1. Size dependence of thermal conductivity of single PE chain ($\kappa_{SPEC}$) at different temperatures (250 ~ 400 K).

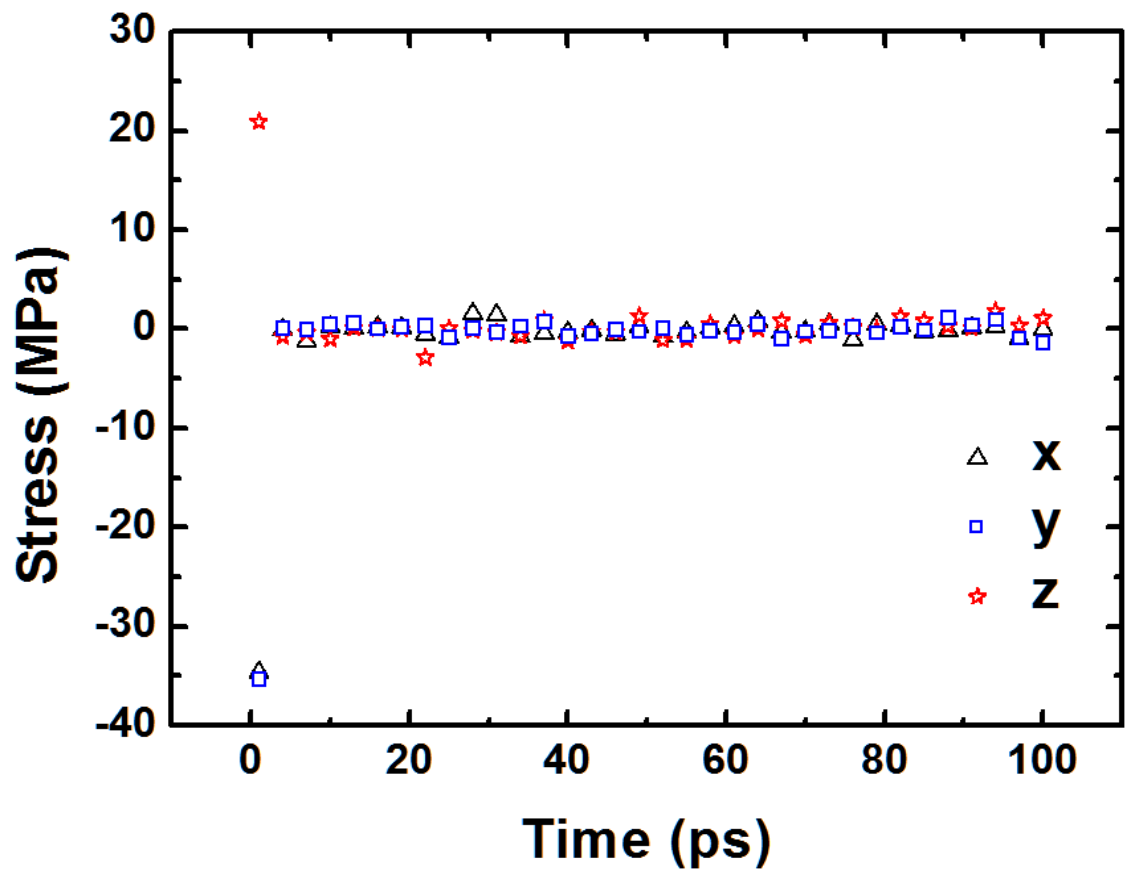

Figure S2. Stress of PPEL system as a function of time during *NPT* relaxation process.